\documentstyle[12pt]{article}
\renewcommand{\thefootnote}{\fnsymbol{footnote}}
\newcommand{\newsection}{
\setcounter{equation}{0}
\section}

\def\appendix#1{
  \addtocounter{section}{1}
  \setcounter{equation}{0}
  \renewcommand{\thesection}{\Alph{section}}
 \section*{Appendix \thesection\protect\indent \parbox[t]{11.715cm} {#1}}
  \addcontentsline{toc}{section}{Appendix \thesection\ \ \ #1}
  }

\newcommand{\tr}[1]{\:{\rm tr}\,#1}
\newcommand{\dd}[1]{\frac{\partial}{\partial{#1}}}
\def\e{{\,\rm e}\,}
\newcommand{\rf}[1]{(\ref{#1})}
\newcommand{\eq}[1]{Eq.~(\ref{#1})}
\newcommand{\non}{\nonumber \\*}
\def\be{\begin{equation}}
\def\ee{\end{equation}}
\def\bea{\begin{eqnarray}}
\def\eea{\end{eqnarray}}

\def\tw{\tilde{W}}

\def\fii{\sqrt{\vphantom{\lambda_j}a\lambda_i+\eta ^2}}
\def\fjj{\sqrt{a\lambda _j+\eta ^2}}

\let\wtd=\widetilde
\let\pa=\partial

\begin{document}
\begin{titlepage}
\begin{flushright}
MIAN-Th-18/98\\
NBI-HE-98-13\\
MPS-RR 1998-8\\
hep-th/9805212\\
\end{flushright}
\vspace{.5cm}

\begin{center}
{\LARGE The  NBI matrix model of IIB Superstrings }\\
\vspace{1.2cm}
{\large J. Ambj{\o}rn${}^{{\rm a)}}$%
\footnote{E-mail: ambjorn@nbivms.nbi.dk}
and L. Chekhov${}^{{\rm b)}}$\footnote{E-mail:
chekhov@genesis.mi.ras.ru} }\\
\vspace{24pt}
${}^{{\rm a)}}${\it The Niels Bohr Institute,} \\
{\it Blegdamsvej 17, 2100, Copenhagen, Denmark}\\
[.2cm]
${}^{{\rm b)}}${\it Steklov Mathematical Institute,}
\\ {\it  Gubkina 8, 117966, GSP--1, Moscow, Russia}
\end{center}
\vskip 0.9 cm
\begin{abstract}
We investigate the NBI matrix model with the potential
$X\Lambda+X^{-1}+(2\eta+1)\log X$ recently
proposed to describe IIB superstrings. With the proper
normalization, using Virasoro constraints, we prove
the equivalence of this model and the Kontsevich matrix model
for $\eta\ne0$ and find the explicit transformation between the two models.
\end{abstract}
\end{titlepage}
\setcounter{page}{1}
\renewcommand{\thefootnote}{\arabic{footnote}}
\setcounter{footnote}{0}

\newsection{Introduction}

The investigation of matrix models with an external field and
a logarithmic potentials was initiated in~\cite{CM,ChZ}. These
matrix models are related to the so-called NBI matrix model,
which appeared recently
in the context of the IIB superstring matrix
model~\cite{BFSS96,IKKT,FMOSZ,KO}. In~\cite{FMOSZ,KO},
the following (M)atrix model action was proposed:
\be\label{jan1}
S_{\hbox{\small NBI}}=-\frac{\alpha}{4}\tr Y^{-1}[A_\mu,A_\nu]^2+
\beta \tr Y+2\eta\tr\log Y
-\frac{1}{2}\tr {\overline\Psi}\Gamma^\mu[A_\mu,\Psi].
\ee
It possesses ${\cal N}\!=\!2$ supersymmetry in the large $N$
limit~\cite{FMOSZ}. The matrix $Y$ plays here the role of the world-sheet
metric to be integrated out in order to obtain the effective action.
Then, the (nonlocal) logarithmic term is a curvature term of the
world-sheet metric. As was shown in~\cite{ChZ}, the relevant choice
of the constant $\eta$ in front of this term  that leads to a
non-Abelian Born--Infeld action for the string coordinates is
$\eta=-\frac{1}{4N}$, $N$ being the matrix size. However, theories
for other values of $\eta$ were also considered~\cite{MMS}, and the
answer for the effective action in the leading order in $N$ was
obtained in~\cite{ChZ} for general $\eta$.

In the present paper, we derive the constraint equations for the NBI
matrix model, and show their coincidence with the
constraint equations for
the Kontsevich matrix model, thereby proving the equivalence of
the two models for nonzero $\eta$.

\newsection{The matrix model in the large $N$ limit}

We start with the following matrix integral:
 \begin{equation}\label{mm}
 Z=\int dX\,\e^{-N\tr\left[X\Lambda +X^{-1}+(2\eta +1)\log X\right]},
 \end{equation}
As shown in \cite{ChZ}, it is related to the bosonic part of \rf{jan1}
by the following change in integration variables: $Y = \frac{N}{\beta} X^{-1}$,
and by
\be\label{jan2}
\Lambda = -\frac{\alpha\beta}{4N^2}\, [A_\mu,A_\nu]^2.
\ee

 The matrix integral \rf{mm} belongs to a class of generalized Kontsevich
 models
(GKM) \cite{GKM}. Such models with negative powers of the matrix~$X$
 have been previously discussed in the context of
 $c=1$ bosonic string theory \cite{DMP}. In \cite{MMS}, the $\tau$-function
 approach to such models was developed. There, the parameter~$\eta$ plays
 the role of the zeroth time in the corresponding integrable hierarchy.
 Moreover, at the conformal point $\eta=0$, this model was shown
 \cite{MMS} to have the same Schwinger--Dyson equations as the $U(N)$
 model solved in \cite{unit,BG}.

 For the models of this type, the
 large $N$ solutions are known explicitly only
 in some special cases.  The models with cubic potential for $X$
 \cite{cubic} and the combination of the logarithmic and quadratic
 potentials \cite{CM} were solved by a method based on Schwinger--Dyson
 equations, developed first for the unitary matrix models with external
 field \cite{unit,BG}.
 The same technique, being applied to the
 integral \rf{mm}, also allows one to find its large $N$ asymptotic
expansions in  the closed form for arbitrary $\eta $~\cite{ChZ}.

  The Schwinger--Dyson equations for \rf{mm} follow from the identity
 \begin{equation}\label{...=0}
 \frac{1}{N^3}\,\frac{\partial }{\partial \Lambda_{jk} }
 \,\frac{\partial }{\partial \Lambda_{li} }\,\int dX\,\,
 \frac{\partial }{\partial X_{ij}}\,
 \e^{-N\tr\left[X\Lambda +X^{-1}+(2\eta +1)\log X\right]}=0 .
 \end{equation}

Written in terms of the eigenvalues, these $N$ equations read
(no summation over $i$ is implied)
 \begin{equation}
\label{SDeig}
 \left[-\frac{1}{N^2}\,\lambda _i\,\frac{\partial ^2}{\partial
 \lambda_i^2} -\frac{1}{N^2}\,\sum_{j\neq i}\lambda _j\,\frac{1}{\lambda
 _j-\lambda _i}\,\left(\frac{\partial}{\partial \lambda _j}-
 \frac{\partial }{\partial \lambda _i}\right)
 +\frac{1}{N}\,(2\eta -1)\,\frac{\partial
 }{\partial \lambda _i}+1\right]Z(\lambda )=0.
 \end{equation}
 For $\eta=0$, these formulas coincide with the
 corresponding formulas for the $U(N)$ model~\cite{unit,BG}.

 It is convenient to set
 \begin{equation}\label{defW}
 W(\lambda _i)=\frac{1}{N}\,\frac{\partial }{\partial \lambda _i}\,\log Z.
 \end{equation}
This quantity plays an important role in evaluating the large~$N$ limit.

 We also introduce the eigenvalue density of the matrix $\Lambda $:
 \begin{equation}\label{dens}
 \rho (x)=\frac{1}{N}\,\sum_{i}\delta (x-\lambda _i).
 \end{equation}
 The density obeys the normalization condition
 \begin{equation}\label{norm}
 \int dx\,\rho (x)=1,
 \end{equation}
 and in the large $N$ limit it becomes a smooth function.

 Simple power counting shows that the derivative of $W(\lambda_i )$ in
 the first term on the left-hand side of Eq. \rf{SDeig} is suppressed
 by the factor $1/N$ and can be omitted at $N=\infty $. The remaining
 terms are as follows:
 \begin{equation}\label{inteqn}
 -xW^2(x)-\int dy\rho (y)\,y\,\frac{W(y)-W(x)}{y-x}+(2\eta -1)W(x)+1=0,
 \end{equation}
 where $\lambda _i$ is replaced by $x$. Equation \rf{inteqn} can be
 simplified by the substitution
 \begin{equation}\label{subst}
 \tw(x)=xW(x)-\eta.
 \end{equation}
 After some transformations, using the normalization condition \rf{norm},
 we obtain
 \begin{equation}\label{maineq}
 \tw^2(x)+x\int dy\rho (y)\,\,\frac{\tw(y)-\tw(x)}{y-x}=x+\eta ^2.
 \end{equation}

 The nonlinear integral equation \rf{maineq} can be solved with the
 help of the ansatz
 \begin{equation}\label{anz}
 \tw(x)=f(x)+\frac{x}{2}\,\int dy\,\frac{\rho
 (y)}{f(y)}\,\frac{f(y)-f(x)}{y-x},
 \end{equation}
 where $f(x)$ is an unknown function to be determined by substituting
 \rf{anz} into \eq{maineq}. The asymptotic behaviors of $\tw(x)$ and
 $f(x)$ as $x\rightarrow \infty $ follow from eq.~\rf{maineq}:
 $\tw(x)\sim \sqrt{x}-1/2$, and the analytic solution with
 minimal set of singularities is
 \begin{equation}\label{fx}
 f(x)=\sqrt{ax+b}.
 \end{equation}
 The parameters $a$ and $b$ are unambiguously determined from
 \eq{maineq}. We find that $b=\eta ^2$ and $a$ is implicitly  defined by
 \begin{equation}\label{defa}
 1+\frac{1}{2}\int dy\,\frac{\rho (y)}{f(y)}=\frac{1}{\sqrt{a}},
 \end{equation}
 or, in terms of the eigenvalues,
 \begin{equation}\label{a1}
 1+\frac{1}{2N}\sum_{j}\frac{1}{\sqrt{a\lambda _j+\eta
 ^2}}=\frac{1}{\sqrt{a}}.
 \end{equation}

Then, the answer for the integral in the large $N$ limit reads~\cite{ChZ}

 \begin{eqnarray}\label{res}
 \log Z&=&N^2\left[\left(\eta ^2+\frac{1}{4}\right)\log a+\frac{4\eta
 ^2}{\sqrt{a}}-\frac{\eta ^2}{a}\right]
 \non  &&
 +N\sum_{i}\left[
 \frac{2}{\sqrt{a}}\,\fii+\eta \log\left(\lambda _i\,\frac{\fii-\eta
 }{\fii+\eta }\right)\right]
 \non  &&
 -\frac{1}{2}\sum_{ij}\log\left(\fii+\fjj\right).
 \end{eqnarray}
 One can verify directly
 that $\frac{\partial }{\partial
 a}\log Z=0$ and $\frac{1}{N}\,\frac{\partial }{\partial \lambda _i}\log
 Z=W(\lambda _i)$, as far as \eq{a1} holds.

\newsection{The Kontsevich phase}

We are interested in the asymptotic expansion of the model (\ref{mm})
for large $\Lambda$. Then, the expansion parameters are traces of
negative powers of the external matrix $\Lambda$. Conventionally,
this regime is called the {\it Kontsevich phase} of the solution.

Here an important note is in order. In~\cite{ChZ}, we did not discuss
which branch of the root---positive or negative---should
be chosen in (\ref{a1}), since both choices led to the same answer
for the integral in the large $N$ limit (\ref{res}).
However, in what follows, we must fix this sign.

The Kontsevich phase is
the strong coupling regime where the expansion in negative
powers of~$\lambda_i$ is to be performed. Then we see that the
dependence is only on $\lambda_i^{-n-1/2}$, \ $n=0,1,\dots$\,.

As the first step,
we perform the phase analysis for the toy case where
all $\lambda_i$'s coincide, so
(\ref{a1}) becomes
\be
\label{a2}
\frac{a}{(1-\sqrt{a})^2}=4(a\lambda+\eta^2),\quad a>0,\ \lambda>0.
\ee
Then, obviously, the sign of the square term in (\ref{a1}) is
negative for $a>1$ and positive for $0<a<1$.

Algbebraically, there always
(except if  $\eta=0$) exist two solutions to (\ref{a2}): one with
$0<a<1$ and another with $a>1$.
For the Kontsevich phase to be possible, we demand that
the expansion in terms of the so-called {\em times}
\be\label{jan3}
\tau_k=\frac{1}{2k-1}\tr \Lambda^{-k+1/2},\quad k=1,2,\dots\,.
\ee
should make sense. So, we assume that
\be
\label{a3}
\left|\frac{\eta^2}{a\lambda}\right|<1.
\ee
Then, from (\ref{a2}) and (\ref{a3}), we obtain the following phase diagram:

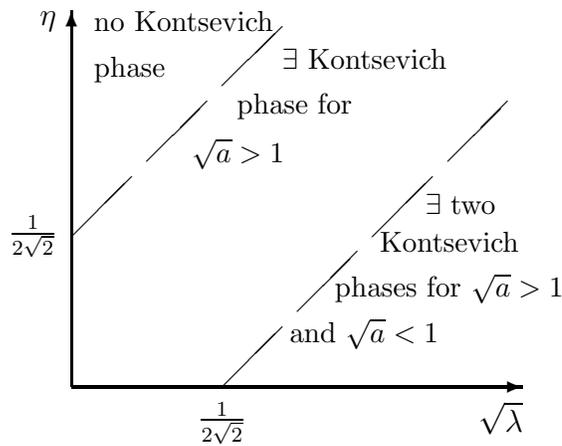
\setlength{\unitlength}{1mm}%
\begin{picture}(50,70)(-40,0)
\thicklines
\put(10,10){\vector(1, 0){60}}
\put(10,10){\vector(0, 1){50}}
\thinlines
\put(10,30){\line(1,1){8}}
\put(20,40){\line(1,1){8}}
\put(30,50){\line(1,1){8}}
\put(30,10){\line(1,1){8}}
\put(40,20){\line(1,1){8}}
\put(50,30){\line(1,1){8}}
\put(60,40){\line(1,1){8}}
\put(13,60){\makebox(0,0)[lt]{\hbox{\small no Kontsevich}}}
\put(13,54){\makebox(0,0)[lt]{\hbox{\small phase}}}
\put(38,55){\makebox(0,0)[lt]{\hbox{\small $\exists$ Kontsevich}}}
\put(32,49){\makebox(0,0)[lt]{\hbox{\small phase for}}}
\put(26,43){\makebox(0,0)[lt]{\hbox{\small $\sqrt{a}>1$}}}
\put(57,33){\makebox(0,0)[lb]{\hbox{\small $\exists$ two}}}
\put(51,28){\makebox(0,0)[lb]{\hbox{\small Kontsevich}}}
\put(45,22){\makebox(0,0)[lb]{\hbox{\small phases for $\sqrt{a}>1$}}}
\put(39,16){\makebox(0,0)[lb]{\hbox{\small and $\sqrt{a}<1$}}}
\put(8,30){\makebox(0,0)[rc]{$\frac{1}{2\sqrt{2}}$}}
\put(30,8){\makebox(0,0)[ct]{$\frac{1}{2\sqrt{2}}$}}
\put(8,60){\makebox(0,0)[rt]{$\eta$}}
\put(70,8){\makebox(0,0)[rt]{$\sqrt{\lambda}$}}
\end{picture}

\newsection{Constraint equations in the Kontsevich phase}

Now we write
(\ref{SDeig}) in terms of the relevant times \rf{jan3}.  Here, to obtain
rigorous results, the normalizing factor is necessary.  From the
theory of the generalized Kontsevich model~\cite{GKM},
the proper expression, which
has no explicit dependence on the matrix size $N$, reads
\be
\label{GKM}
{\cal Z}(\{\tau_n\})=\frac{\int DX\e^{\Lambda X+V(X)}}{
\e^{\tr \Lambda X_0+V(X_0)}\det^{-1/2}\left(
\frac{\delta}{\delta X}\otimes \frac{\delta}{\delta X}V(X_0)\right)}.
\ee
Here $X_0$ is the stationary point, $\Lambda+V'(X_0)=0$, and
the determinant in the normalizing factor comes from the
quasi-classical integration.

Let us choose the new variables
\be
\label{ZvisL}
\lambda_i=z_i^2-(\eta+1/2)^2.
\ee
Then the normalizing factor reads
\be
\label{Norm}
\exp\left\{-N\sum_{i}^{}\bigl[2z_i-2\eta\log(z_i+\eta+1/2)\bigr]
-\frac{1}{2}\sum_{i,j}^{}\log(z_i+z_j)\right\}.
\ee

Note that if we perform the standard Itzykson--Zuber integration,
expression (\ref{GKM}) for integral (\ref{mm}) becomes
\be
{\cal Z}(\{\tau_n\})=\frac{\mathop{\det}\limits_{{1\le i\le N\atop 0\le l\le
N-1}}\left|\left| \xi_i^{2\eta N-l}K_{-2\eta N-l}(N\xi_i)\right|\right|}
{\prod_i \e^{-2Nz_i}(z_i+\eta+1/2)^{2\eta N}(2z_i)^{-1/2}\Delta(z)},
\qquad \xi_i\equiv \sqrt{\lambda_i},
\label{detK}
\ee
where $K_\nu(x)$ are Macdonald functions,
$K_\nu(x)=\int_{-\infty}^{\infty}ds\,\e^{-2x\cosh s+\nu s}$,
and (\ref{detK}) does not resemble too much the corresponding expression
for the Kontsevich matrix model where these Airy functions stand instead
of $K_\nu(x)$. To find the large $N$ asymptotic behavior of (\ref{detK}),
we use the constraint equation method.

In terms of $z$-variables, (\ref{SDeig}) becomes
\bea
\label{SDz}
&{}&\left[-\frac{1}{N^2}\bigl(z_i^2-(\eta+1/2)^2\bigr)\frac{1}{2z_i}
\dd{z_i}\left(\frac{1}{2z_i}\dd{z_i}\right)\right.\nonumber\\
&{}&\left.-\frac{1}{N^2}\sum_{j\ne i}^{}
\frac{z_j^2-(\eta+1/2)^2}{z_j^2-z_i^2} \left(\frac{1}{2z_j}\dd{z_j}
-\frac{1}{2z_i}\dd{z_i}\right)+\frac{2\eta-1}{N}\frac{1}{2z_i}\dd{z_i}
+1\right]\,Z(z_i)=0
\eea
When pushing the normalizing factor (\ref{Norm}) through derivatives
w.r.t.\ $z_j$-variables, we replace
$\left(\partial_i\equiv \dd{z_i}\right)$
\be
\label{repl}
\partial_i\to \partial_i-2N+2\eta N\frac{1}{z_i+\eta+1/2}
-\sum_{j}^{}\frac{1}{z_i+z_j}.
\ee

Let us introduce the new {\it times}
\be
\label{times}
t_{n+1}=\frac{1}{2n-1}\sum_{i}^{} \frac{1}{z_i^{2n-1}}
+\delta_{n,0}\frac{N}{\eta+1/2}, \qquad n=0,1,\dots,
\ee
which differs slightly from the conventional ones defines above by \rf{jan3}.
As is easily checked, they are related by a lower triangular transformation,
so they are equivalent from the view point of phase transitions
and critical behavior.

Then, the constraint equations for ${\cal Z}(\{t_n\})$
are obtained after some tedious algebra which we omit here. Collecting
all coefficients to the term
$\frac{1}{z_i^{2s+4}}\equiv \dd{z_i}t_{s+2}$,
we obtain
\be
\label{wtdLs}
\wtd L_s{\cal Z}(\{t_n\})=0, \qquad s\ge-1,
\ee
where
\bea
\wtd L_s&=&\delta_{s,-1}\left[-\frac{1}{16N^2}
+t_1^2\frac{(\eta+1/2)^2}{4N^2}\right]
+\delta_{s,0}\frac{(\eta+1/2)^2}{16N^2}\nonumber\\
&{}&\qquad-\frac{1}{2N^2}\sum_{{p=0\atop p+s\ge0}}^{\infty}
\left[-(\eta+1/2)^2(2p+1)t_{p+1}+(2p-1)(1-\delta_{p,0})t_p\right]
\frac{\pa}{\pa t_{p+s+1}}\nonumber\\
&{}&\qquad-\frac{1}{N}\left(\frac{\pa}{\pa t_{s+2}}-\frac{\eta+1/2}{2}
(1-\delta_{s,-1})\frac{\pa}{\pa t_{s+1}}\right)\nonumber\\
&{}&\qquad-\frac{1}{4N^2}\sum_{k=1}^{s+1}\frac{\pa}{\pa t_k}
\frac{\pa}{\pa t_{s+2-k}}+\frac{(\eta+1/2)^2}{4N^2}
\sum_{k=1}^{s}\frac{\pa}{\pa t_k}\frac{\pa}{\pa t_{s+1-k}}.
\label{Lss}
\eea
In the convenient conformal field theory notation, (\ref{Lss}) becomes
\be
\label{Ls}
\wtd L_s=V_{s+1}-(\eta+1/2)^2V_s
+\frac{4\eta N}{\eta+1/2}\alpha_{2s+3}.
\ee
where
\bea
\label{alpha+}
&{}&\Bigl.\alpha_{2n+1}\Bigr|_{n\ge 0}\equiv \dd{t_{n+1}},\\
\label{alpha-}
&{}&\Bigl.\alpha_{-2n-1}\Bigr|_{n\ge 0}\equiv (2n+1)t_{n+1},\\
\label{comm}
&{}&[\alpha_{-2k-1},\alpha_{2q+1}]=-(2k+1)\delta_{k,q},
\eea
and the operators $V_s$ in the free-field
representation are quadratic in $\alpha_i$,
\be
\label{Vs}
V_q\equiv \sum_{a,b}^{}\delta_{q, a+b+1}
{:}\alpha_{2a+1}\alpha_{2b+1}{:}+1/4\delta_{s,0}\,,
\ee
where the normal ordering ${:}\ {:}$ means that all $\alpha_a$ with positive
indices are on the right of all $\alpha_b$ with negative indices.

For $s,t\ge-1$, \ $\wtd L_s$ satisfy the algebra
\be
[\wtd L_s, \wtd L_t]=4(s-t)\bigl(\wtd L_{s+t+1}
-(\eta+1/2)^2\wtd L_{s+t}\bigr),
\ee
from which the Virasoro algebra can be obtained by the lower-triangle
replacement
\be
\label{replac}
L_s\equiv \wtd L_s+\sum_{k=1}^{\infty}(\eta+1/2)^{-2k}\wtd L_{s+k}.
\ee
Performing replacement (\ref{replac}) and rescaling $V_s$,
we obtain the Virasoro algebra in terms of the generators
\be
\label{11}
L_s=-(\eta+1/2)^2V_s+4\eta
N\sum_{k=2}^{\infty}\frac{1}{(\eta+1/2)^{2k-3}} \partial_{s+k}.
\ee
Amazingly, if we manage to remove the derivative terms in (\ref{11}),
then the constraints we obtain will be just constraints of the
Kontsevich matrix model~\cite{MS,IZ92}.
To remove these terms,
we shift all of
the higher times, leaving the times $t_1$ and $t_2$
unshifted,
\be
\label{timeshift}
\tilde t_k\equiv t_k-\frac{4\eta N}{(\eta+1/2)^{2k+1}}
\frac{1}{2(2k+1)}, \quad k\ge3, \qquad \tilde t_{1,2}=t_{1,2}.
\ee
Explicitly, in terms of the new times,
\bea
\label{L-1}
&{}&L_{-1}=\tilde t_1^2+2\sum_{m=1}^{\infty}(2m+1)\tilde t_{m+1}
\dd{\tilde t_m}-\frac{4\eta N}{(\eta+1/2)^3}\partial_{1}\\
\label{Lss1}
&{}&L_{s}=\sum_{m=1}^{s}\dd{\tilde t_m}\dd{\tilde t_{s-m}}
+2\sum_{m=1}^{\infty}(2m-1)\tilde t_{m}
\dd{\tilde t_{m+s}}+\frac{1}{4}\delta_{s,0}
-\frac{4\eta N}{(\eta+1/2)^3}\dd{\tilde t_{s+2}}.
\eea
This is nothing but the Virasoro algebra that appears in the Kontsevich
matrix model. Therefore, we have proven the equivalence between the model
(\ref{mm}) and the Kontsevich matrix model~\cite{Kon}.

\newsection{A large $N$ limits comparison}

Let us explicitly compare
the model (\ref{mm})
after the time changing (\ref{timeshift})
and the Kontsevich model with the partition function
\be
\label{KonMod}
Z(M)=\frac{\int dX\exp-\tr\,\gamma\left\{MX^2/2-iX^3/6\right\}}
{\int dX\exp-\tr\,\gamma MX^2/2},
\quad M=\hbox{\,diag\,}\{m_1,\dots,m_N\}.
\ee
Let us consider the constraint equation (\ref{a1}). We have
\be
\label{svyaz}
\frac{1}{2n-1}\tr{M^{-2n+1}}=\tilde t_n=
\frac{1}{2n-1}\tr{Z^{-2n+1}}+\left\{
\begin{array}{ll}
\frac{N}{\eta+1/2},&n=1,\\
0,&n=2,\\
-\frac{1}{2(2n-1)}\frac{4\eta N}{(\eta+1/2)^{2n-1}},&n\ge3,
\end{array}
\right.
\ee
i.e.,
\be
\label{MZ}
\tr M^{-2n+1}=\tr Z^{-2n+1}-\frac{1}{2}\frac{4\eta N}{(\eta+1/2)^{2n-1}}
+\frac{1}{2}\frac{4\eta N}{(\eta+1/2)^3} \delta_{n,2}+2N\delta_{n,1}.
\ee
Then we obtain, in terms of these shifted
variables, ($\pm$ depends on the branch of the square root),
\bea
\frac{1}{\sqrt{a}}&=&1 \pm\frac{1}{2N}\sum_{j}^{}
\frac{1}{\sqrt{az_j^2-a(\eta+1/2)^2+\eta^2}}\nonumber\\
&=&1\pm \frac{1}{2N\sqrt{a}}\sum_{j}^{}\sum_{n=0}^{\infty}
\frac{(2n-1)!!}{(2n)!!}\frac{1}{z_j^{2n+1}}
\bigl((\eta+1/2)^2-\eta^2/a\bigr)^n\nonumber\\
&=&1\pm \frac{1}{2N\sqrt{a}}\left[
\sum_{j}^{}\frac{1}{\sqrt{m_j^2+\eta^2/a-(\eta+1/2)^2}}\right.
\nonumber\\
&&\qquad\left.+2N\sqrt{a}-2N
-\frac{\eta N}{\eta+1/2}+\frac{\eta^3
N}{(\eta+1/2)^3a} \right].  \label{m1}
\eea
Assuming the minus sign
and denoting
\be
\label{s-gamma}
(\eta+1/2)^2-\eta^2/a\equiv 2s,
\qquad -\frac{2\eta N}{(\eta+1/2)^3}\equiv \gamma,
\ee
we obtain
\be
\label{KontMom}
\sum_{i=1}^{N}\frac{1}{\sqrt{m_i^2-2s}}=-\gamma s,
\ee
i.e., the constraint equation of the Kontsevich model
itself~\cite{MS,IZ92}.

Also, let us recall the answer in the large $N$ limit for the
Kontsevich model~\cite{MS}. If we reconstruct the dependence
on the coupling constant $\gamma$, then
it reads (in the original notation)
\bea
F_0&=&\gamma^2\frac{s^3}{6}+\frac{\gamma}{3}\sum_{i=1}^{N}
\left\{m_i^3-(m_i^2-2s)^{3/2}-3s\sqrt{m_i^2-2s}\right\}
\nonumber\\
&{}&\qquad -\frac{1}{2}\sum_{i,j=1}^{N}
\log\frac{\sqrt{m_i^2-2s}+\sqrt{m_j^2-2s}}{m_i+m_j},
\label{Kon-1}
\eea
where $s$ is determined from
the constraint equation (\ref{KontMom}).

Let us compare the answer (\ref{Kon-1}) with formula (\ref{res})
while accounting for the normalizing condition (\ref{Norm}). Then,
in variables $z_i$, assuming the minus sign in
the constraint (\ref{a1}), we have
\bea
\wtd F_0
&=&N^2\eta^2\log a+\frac{4N^2\eta^2}{\sqrt{a}}-\frac{N^2\eta^2}{a}
\nonumber\\
&+&N\sum_{i}^{}2\left(z_i-\sqrt{z_i^2-2s}\right)
+N\eta\sum_{i}^{}\log\left(\frac{z_i^2-(\eta+1/2)^2}
{\left(z_i+\eta+1/2\right)^2}\right)\left(
\frac{-\sqrt{z_i^2-2s}-\eta/\sqrt{a}}{-\sqrt{z_i^2-2s}+\eta/\sqrt{a}}\right)
\nonumber\\
&-&\frac{1}{2}\sum_{i,j}^{}\log
\left(\frac{\sqrt{z_i^2-2s}+\sqrt{z_j^2-2s}}{z_i+z_j}\right).
\label{resZ}
\eea

After a tedious algebra, taking into account (\ref{MZ}),
we obtain
\be
\label{comp}
\wtd F_0=F_0+2N^2\eta^2\log\frac{\eta}{\eta+1/2}
+N^2\left[\frac{7}{3}\eta^2+\eta-\frac{1}{4}\right],
\ee
where $F_0$ is given by (\ref{Kon-1}) with the substitution (\ref{s-gamma}).
The difference between $\wtd F_0$ and $F_0$ is just
irrelevant constant terms. This again proves that the
two matrix models under consideration coincide.

\newsection{Higher genus expressions}

In this section, we set the Kontsevich coupling constant $\gamma=1$.
The higher genus contributions in the Kontsevich model
are expressed in terms of the so-called moments $I_k$,
\be
I_0=-\frac{1}{N}\sum_{i=1}^{N}\frac{1}{\sqrt{m_i^2-2s}},\qquad
I_k=\frac{\pa^k}{\pa s^k}I_0,\quad k>0.
\ee
Then, for $g>1$, we have~\cite{IZ92}
\be
F_g=\sum_{\sum_{k=2}^{}(k-1)l_k=3g-3}^{}
\langle\tau_2^{l_2}\tau_3^{l_3}\cdots\tau_{3g-2}^{l_{3g-2}}\rangle_g
\frac{1}{(1-I_1)^{2(g-1)+\sum_{}^{}l_p}}
\frac{I_2^{l_2}}{l_2!}\frac{I_3^{l_3}}{l_3!}\cdots
\frac{I_{3g-2}^{l_{3g-2}}}{l_{3g-2}!},
\label{IZ_g}
\ee
i.e., $F_g$ is a finite sum of monomials in $I_k/(1-I_1)^{(2k+1)/3}$ with
coefficients being the intersection indices on the moduli space~\cite{Kon}.
For $g=1$, $F_1=\frac{1}{24}\log\frac{1}{1-I_1}$.

Now we rewrite the expression (\ref{IZ_g}) via the moments of the model
(\ref{mm}). Let us introduce the new moments $J_k$
\be
J_k=-(2k-1)!!\frac{1}{N}\sum_{j=1}^{N}\frac{1}{(a\lambda_j+\eta^2)^{k+1/2}},
\quad k=0,1,2,\dots\,.
\label{moment}
\ee
We are interested only in transformation law for $I_k$ with $k>0$
since the only dependence on the moment $I_0$ is via the constraint
equation (\ref{KontMom}). Then, Eq.~(\ref{MZ}) implies
\bea
&{}&(I_1-1)\to a^{3/2}(J_1+2/\eta^2),\nonumber\\
&{}&I_k\to a^{k+1/2}(J_k+2/\eta^{2k}),\nonumber
\eea
i.e., the expression (\ref{IZ_g}) becomes
\be
F^{NBI}_g=\sum_{\sum_{k=2}^{}(k-1)l_k=3g-3}^{}
\langle\tau_2^{l_2}\tau_3^{l_3}\cdots\tau_{3g-2}^{l_{3g-2}}\rangle_g
\frac{1}{\bigl(J_1+\frac{2}{\eta^2}\bigr)^{2(g-1)+\sum_{}^{}l_p}}
\prod_{k=2}^{3g-2}
\frac{\bigl(J_k+\frac{2}{\eta^{2k}}\bigr)^{l_k}}{l_k!},\quad g>1
\label{mm_g}
\ee
and
\be
F^{NBI}_1=\frac{1}{24}\log\bigl[a^{3/2}(J_1+2/\eta^{2})\bigr].
\label{mm_1}
\ee

Therefore, expression (\ref{res}) for genus zero, taking into account
the normalizing factor (\ref{Norm}) and the expressions (\ref{mm_1}),
(\ref{mm_g}) completely determine the partition function of the
model (\ref{mm}) for all genera. These expansions are, however,
ill-defined for $\eta\to 0$, which corresponds to the $U(N)$ model,
and for $\eta\sim 1/N$, which corresponds to the model defined
in~\cite{FMOSZ}.

\newsection{Remarks}

{\bf 1.}
The last observation above is related to the initial constraints (\ref{wtdLs}).
Note that we can consider a ``minimal reduction,'' where all times
$t_k$ but the time $t_1$ are equal to zero. Then this partition function
is entirely determined from the action of $\wtd L_{-1}$ on
$\cal Z$,
$$
\left.\left(\frac{4\eta }{\eta+1/2}+2t_1\right)
\frac{\partial{\cal F}}{\partial t_1}\right|_{t_2=t_3=\dots=0}
-N^2(\eta+1/2)^2t_1^2+1/4=0,
$$
where
$$
t_1\equiv\frac{1}{N}\sum_{i=1}^{N}z_i^{-1}+(\eta+1/2)^{-1}
\bigl|_{t_2=t_3=\dots=0}
=\frac{1}{N}\sum_{i=1}^{N}\lambda_i^{-1/2}+(\eta+1/2)^{-1},
$$
i.e.,
\be
\label{Ft1}
\left.\frac{\partial{\cal F}}{\partial t_1}\right|_{t_2=t_3=\dots=0}
=\frac{\eta+1/2}{4}\left[N^2\bigl((\eta+1/2)t_1-2\eta\bigr) +
\frac{(2\eta N)^2-1/4}{(\eta+1/2)t_1+2\eta }\right].
\label{551}
\ee
Remark that ${\cal F}$ becomes a polynomial in $t_1$ for
$$
\eta=\pm \frac{1}{4N},
$$
and this is exactly the point corresponding to the IIB superstring
model~\cite{FMOSZ}.

On the other hand,
from expressions (\ref{mm_g}), (\ref{mm_1})
it is clear that
in this case all $J_k=0$ for $k>0$, so there is no $t_1$-dependent
contributions from $g>1$ and only $\frac{1}{16}\log a$ comes from
the genus one term.
One can compare with expression for $F_1$ coming from (\ref{551}) and,
taking into account the constraint equation (\ref{a1}),
one finds an exact coincidence
of the two expressions, i.e.,
$$
F_1|_{t_2=t_3=\dots=0}=-1/8\log(2+1/N\tr
\lambda^{-1/2})=\frac{1}{16}\log a.
$$

{\bf 2.}
The string susceptibility in the large $N$ limit
can be obtained by differentiating twice w.r.t.\
the string coupling constant $\eta$. One should check that after the
first differentiation of (\ref{res}),
the stationary condition
still holds, so the total
derivative coincides with the partial one and, as a result, we have
\be
\label{dda}
\frac{\hbox{d}^2\log Z}{\hbox{d}\eta^2}=
\frac{\partial^2\log Z}{\partial\eta^2}=2N^2(\log a+3).
\ee

\section*{Acknowledgement}
J.~A. acknowledges the support of the Danish National
Research Foundation.
L.~Ch. is grateful to the Niels Bohr Institute for the hospitality. The
work of L.~Ch. was supported in part by the Russian Foundation for Basic
Research, Grant No.~96-01-00344.

\end{document}